\documentclass{emulateapj}
\usepackage{apjfonts}
\usepackage{graphicx}
\usepackage[dvips]{color}
\usepackage{amsmath}
\usepackage{amsfonts}
\usepackage{amssymb}
\usepackage[letterpaper=true,hyperfootnotes=false,colorlinks=true]{hyperref}

\newcommand{\as}{$\arcsec$}
\slugcomment{Draft, Submitted to ApJ.}
\shorttitle{Blue Luminescence and Small PAHs in the ISM}
\shortauthors{Vijh et al.}

\begin{document}

\title{Blue Luminescence and the Presence of Small PAHs in the ISM}

\author{Uma P. Vijh\altaffilmark{1,2}, Adolf N. Witt\altaffilmark{1,2} \& Karl D. Gordon\altaffilmark{3}}

\altaffiltext{1}{Visiting Astronomer, Cerro Tololo Inter-American Observatory.CTIO is operated by AURA, Inc.\ under contract to the National Science Foundation.}
\altaffiltext{2}{Ritter Astrophysical Research Center, The University of Toledo,Toledo, OH 43606, (uvijh@astro.utoledo.edu, awitt@dusty.astro.utoledo.edu)}
\altaffiltext{3}{Steward Observatory, University of Arizona, Tucson, AZ 85721, (kgordon@as.arizona.edu)}

\begin{abstract}
Blue Luminescence (BL) was first discovered in a proto-planetary nebula, the Red Rectangle (RR) surrounding the post-AGB star HD~44179. BL has been attributed to fluorescence by small, 3-4 ringed neutral polycyclic aromatic hydrocarbon (PAH) molecules, and was thought to be unique to the RR environment where such small molecules are actively being produced and shielded from the harsh interstellar radiation by a dense circumstellar disk. In this paper we present the BL spectrum detected in several ordinary reflection nebulae illuminated by stars having temperatures between 10,000 -- 23,000~K.  All these nebulae are known to also exhibit the infrared emission features called aromatic emission features (AEFs) attributed to large PAHs. We present the spatial distribution of the BL in these nebulae. In the case of Ced~112, the BL is spatially correlated with mid-IR emission structures attributed to AEFs. These observations provide evidence for grain processing and possibly for in-situ formation of small grains and large molecules from larger aggregates. Most importantly, the detection of BL in these ordinary reflection nebulae suggests that the BL carrier is an ubiquitous component of the ISM and is not restricted to the particular environment of the RR.
\end{abstract}


\section{INTRODUCTION}
We discovered the presence of small, 3-4 ringed, neutral polycyclic aromatic hydrocarbons (PAHs) in the Red Rectangle (RR) through their electronic fluorescence, called Blue Luminescence (BL) ($\lambda_{peak}\sim3750$~\AA) \citep{vijh04,vijh05}. The observed spectrum was attributed to small PAHs on the basis of a comparison with laboratory spectra of PAH molecules fluorescing in the gas phase \citep{vijh04}, the spatial correlation between the BL and the 3.3 $\micron$ emission, commonly attributed to small, neutral PAH molecules, and the newly-derived UV/optical attenuation curve for the central source of the RR, HD~44179 \citep{vijh05}. The latter two results provided strong independent evidence for the presence of small PAH molecules with masses of less than 250 amu in the RR, which supports the attribution of the blue luminescence to fluorescence by the same molecules. Such fluorescence is excited by the absorption of near-ultraviolet photons, followed by the relaxation to the lowest level of the first excited electronic state via a series of mid-IR vibrational transitions. The subsequent \emph{electronic} transitions to various vibrational levels of the electronic ground state are then observed as optical fluorescence, i.e. the BL. The BL is much more specific in terms of size and ionization state of the emitter than vibrational transitions observed in the mid-IR \citep{reyle00}. In particular, the wavelength of the most energetic electronic transition seen in neutral PAHs is closely dependent on the size of the molecular species. In general, this fluorescence wavelength increases with the molecular weight and size of the molecule \citep{vijh04}.

 Other emission bands that are most commonly attributed to PAHs are the family of features at 3.3, 6.2, 7.7, 8.6, 11.3, \& 12.7 $\mu$m (referred to as aromatic emission features (AEF) or unidentified infra-red (UIR) bands) \citep{hony01,cook98,lp84,ver01,atb85,sel84}. These AEFs are the signatures of aromatic C-C and C-H fundamental vibrational and bending modes and the specific sizes  and ionization states of the PAH molecules or ions cannot be ascertained from these spectra alone. On absorption of a far-UV photon, a PAH molecule or ion usually undergoes a transition to an upper electronic state. If the molecule or ion  undergoes iso-energetic transitions to highly vibrationally excited levels of the \emph{ground state}, then the molecule or ion relaxes entirely through a series of IR emissions in the C-C and C-H vibrational and bending modes. These transitions are largely independent of size, structure and ionization state of the molecule and are identified with the AEFs. The AEFs are found in almost all astrophysical environments including the diffuse interstellar medium (ISM), the edges of molecular clouds, reflection nebulae, young stellar objects, HII regions, star forming regions, some C-rich Wolf-Rayet stars, post-AGB stars, planetary nebulae, novae, normal galaxies, starburst galaxies, most ultra-luminous infra-red galaxies and AGNs \citep[see][and references therein]{pet04}. The presence of the AEF carriers in such a wide range of environments suggests that they must include a wide range of sizes of aromatic structures in several ionization states. Therefore, an observation of BL in astronomical sources in the UV/visible range is of particular value because it offers the possibility of tracing the specific presence of small, most likely neutral PAHs in the ISM. 

The central star in the RR (where the BL was first discovered), HD44179, is a post-AGB star that is in an actively dust-producing stage. Small dust grains and PAH molecules in a wide range of sizes are currently being produced in its circum-stellar outflow environment. The discovery of BL in such an environment was therefore not a surprise. However, several detailed investigations \citep[][and references therein]{jbl99,lepage03}\citep[see also][]{allain96a,allain96b} have led to the conclusion that PAH molecules with fewer than 30 carbon atoms, i.e. the ones detected in the RR,  do not survive in the interstellar environment, once they are ejected from their place of formation. Photo-dissociation by the interstellar radiation field is expected to limit their life time severely.  However, as mentioned earlier, the AEFs are seen in almost all astrophysical environments including those with harsh interstellar radiation fields. Do the small PAHs that produce the BL survive under interstellar conditions or are the AEFs primarily due to larger PAHs? The presence or absence of BL in environments with differing levels of UV radiation can help answer these questions. Presence of BL preferentially traces the  small, neutral PAH molecules, as the larger molecules, even if they were neutral, have generally lower fluorescence efficiencies, and ionized PAH molecules do not fluoresce at optical wavelengths. Small, partially dehydrogenated PAH radicals or radical ions having a singlet ground state, however, could contribute to the BL as well (W.W. Duley, Private communication). To test the viability of such small PAH structures in the general ISM far from C-rich stellar outflows, we obtained long-slit spectra of several reflection nebulae and star-forming regions where AEFs and other emission features like the extended red emission (ERE) have been detected. The temperatures of the illuminating stars in these regions cover a range of 8250 -- 23000~K, and thus offer a chance to study the distribution small PAH molecules as a function of environment.

\section{OBSERVATIONS}
The objects observed for this investigation are listed in Table~\ref{tab-obs}. The observations of the nebula Ced~201 were made at the Steward Observatory 2.3-m Bok telescope on 2002 November 5 and those for NGC~2023, NGC~5367 and Ced~112 were made at the Cerro Tololo Inter-American Observatory (CTIO) 1.5-m telescope on 2003 March 26 and 28. At the Bok telescope we used the Boller and Chivens spectrograph with a grating having 300~lines~mm$^{-1}$, blazed at 376~nm in the first order with a L3800 cut-on filter. The detector used was a 1200~$\times$~800 CCD (ccd20) yielding a spatial scale of 0$\farcs$8 pixel $^{-1}$. The slit used was 4 $\arcmin$ long and 4$\farcs$5 wide. At the CTIO 1.5-m we used the R-C spectrograph with the grating \#09 having 300~lines~mm$^{-1}$, blazed at 400~nm in the first order with a CuSO$_4$ filter. The detector used was the Loral IK CCD and the spectral coverage was from 340~nm to 587.5~nm. The slit used was 7 $\arcmin$ long and 2$\farcs$5 wide and yielded a spatial scale of 1$\farcs$3 pixel $^{-1}$. A coronographic decker assembly was used to minimize scattered light from the star while probing as much of the inner nebula as possible at CTIO. For all observations the nebular exposures were bracketed by exposures of the illuminating stars and the long-slit enabled simultaneous exposures of the nebula and the sky. Table~\ref{tab-obs} gives the details of exposure times and slit positions. Data reductions were carried out with IRAF 2.12 EXPORT, and all spectra were flux-calibrated via observation of standard stars.

%place table of observations here
\begin{deluxetable*}{lllcccc}
\tabletypesize{\scriptsize}
\tablecaption{Details of objects observed, their illuminating stars and slits used.\label{tab-obs}}
\tablewidth{0pt}
\tablehead{
\colhead{Object} & \colhead{Observatory} & \multicolumn{3}{c} {Illuminating Star} & \colhead{Slit Positions (Offset)} & \colhead{Exposures (s)}\\
\cline{3-5}\\
\colhead{} & \colhead{} & \colhead{Name} & \colhead{Spectral Type} & \colhead{Temperature (K)} & \colhead{} & \colhead{}
}
\startdata
Ced~201 RN & Bok 2.3 m & BD~+69~1231 & B9.5 V & 10,000 &  15$\arcsec$north, 20$\arcsec$ south & 4$\times$1800 \\
Ced~112 RN & CTIO 1.5 m & HD~97300 & B9 V & 10,500 & 0$\arcsec$,6$\arcsec$ north, 6$\arcsec$ south& 2$\times$ 1800\\
Ced~112 RN & CTIO 1.5 m & HD~97300 & B9 V & 10,500 & 10 $\arcsec$ north & 3$\times$ 1800\\
NGC~5367 RN & CTIO 1.5 m & Her~4636 & Herbig Ae/Be & 18,700 & 5$\arcsec$ north & 2$\times$1800\\
NGC~2023 RN, PDR & CTIO 1.5 m & HD~37903 & B1.5V & 23,000 & 0$\arcsec$ & 2$\times$1800\\
\enddata
\end{deluxetable*}

\section{RESULTS}
\subsection{Ced 201}
Ced~201 is a rather compact object at a distance of 420~pc \citep{casey91}, on the edge of a molecular cloud. It is excited by the B9.5V star BD~+69~1231. \citet{witt87} note that the radial velocity of this star differs from that of the molecular cloud by 11.7$\pm$3.0~km~s$^{-1}$ so that Ced~201 is probably the result of an accidental encounter of the star with the molecular cloud, while for most other reflection nebulae the exciting star was born in situ. An arc-like structure located between the star and the denser parts of the cloud to the north and the north-east at about 18$\arcsec$ from the star may represent a shock due to the supersonic motion of the star. The mid-IR ISO spectra of Ced~201 give evidence for transformation of very small carbonaceous grains into the carriers of the AEFs, due to the radiation field of the illuminating star and/or to shock waves created by its motion \citep{cesarsky00}. Ced~201 exhibits ERE \citep{wb90} and AEFs \citep{cesarsky00}.

%place Ced201-slit
\begin{figure}
\includegraphics[width=3in]{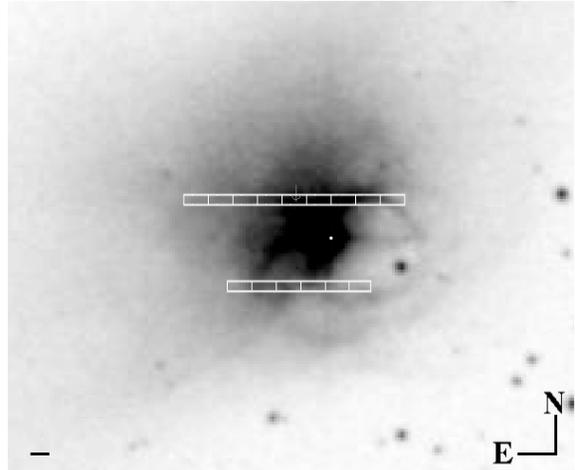}
\caption{DSS.J.POSSII image of Ced~201, overlaid with our slits showing the extracted apertures. The while central pixel indicates the position of the star BD~+69~1231. The dash at the bottom left is 10\as. The right-most extraction window on the northern slit captures the spectrum of the arc-like structure seen on this image.}
\label{Ced201-slit}
\end{figure}

\begin{figure}
\includegraphics[width=3.3in]{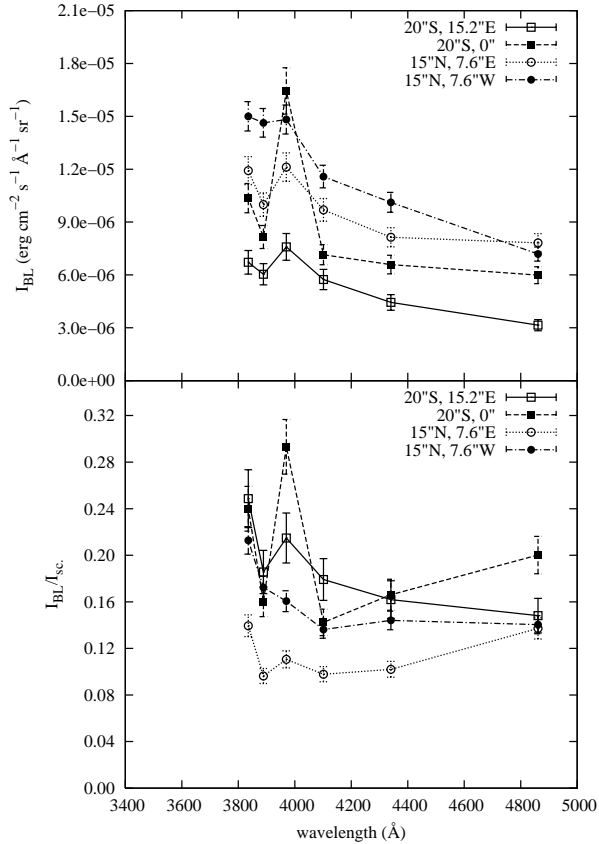}
\caption{BL spectra (upper panel) and ratio of BL to the continuum (lower panel) at 4 different positions in Ced~201.}
\label{Ced201figs}
\end{figure}

We obtained nebular spectra along two slits (PA~90$^\circ$), 15$\arcsec$ north and 20\as\ south of the central star. Figure~\ref{Ced201-slit} shows an optical image of the nebula overlaid with the extracted apertures. We extracted 9 spectra along the 15\as\ north slit and 6 spectra along the 20\as\ south slit each covering 68.4 sq. arcsec of the nebula. BL was measured using the line-depth technique \citep{vijh04} at each of the six hydrogen Balmer lines from H$\beta$ to H$\eta$ in each of these apertures. The line-depth technique detects the BL intensity at the wavelength positions of strong absorption lines by virtue of the fact that the presence of BL reduces the line depths in the dust-scattered nebular spectrum compared to the depths of the identical lines in the spectrum of the illuminating star.  Figure~\ref{Ced201figs} show representative BL spectra in four such apertures and the ratio of BL to the continuum in the same apertures. The use of a cut-on filter at 3800~\AA\ restricted the measurement of the last Balmer line measured to 3835~\AA. Longward of 3800~\AA\ these spectra are similar to the BL spectra seen in the  RR \citep{vijh04,vijh05} and other nebulae discussed in further sections of this paper. The spectra from different offsets do differ, as was also noted by \citet{vijh05} and can probably be attributed to differing radiation environments. Given the poor spectral resolution of BL spectra measured using the line-depth technique, in this paper, we will not characterize the differences in the spectra but will simply refer to them as BL spectra. 

%place Ced 201 figs

\subsection{Ced 112}
Ced~112 is a prominent reflection nebula in the cloud Chamelion I at a distance of $\sim$~160~pc \citep{whittet97}. It is illuminated by the pre-main sequence Herbig AeBe star HD~97300, spectral type B9 \citep{rydgren80} with a luminosity of $\sim$~35~L$_\sun$ \citep{ancker97}. Its pre-main sequence evolutionary status as a Herbig AeBe star is based on indirect clues, namely the association with a reflection nebulosity, the presence of an infrared excess at  $\lambda >$ 5 $\mathrm{\mu}$m, and its location on the ZAMS in the HR diagram \citep{whittet97}. There is no evidence of significant infrared excess at wavelengths shorter than  5~$\micron$ nor of H$\mathrm{\alpha}$ emission. HD~97300 is very likely a relatively old object among HAeBe stars, representative of the latest stages of the pre-main sequence evolution. There is general agreement that the star is an embedded member of the young stellar population and is thus situated at a distance coincident with that of the cloud itself. Infrared emission from the nebula Ced~112 is dominated by AEFs centered at 6.2, 7.7, 8.6, 11.3, and 12.7 $\mathrm{\mu}$m with very small contributions from continuum emission at longer wavelengths \citep{siebenmorgen98}. \citet{siebenmorgen98} also report the detection of an extended, ring-like structure around HD~97300, whose emission is dominated by the AEFs. 

%place Ced112-slits
\begin{figure}
\includegraphics[width=3.3in]{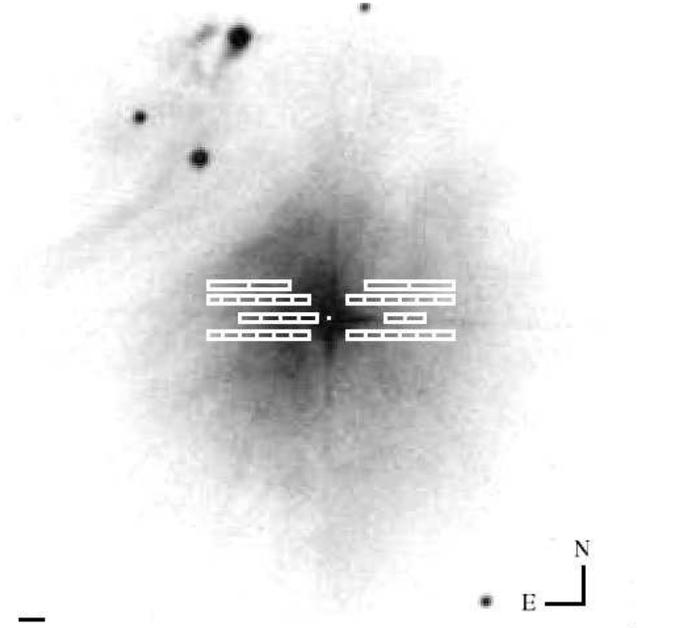}
\caption{AAO.R.DSS2.038 image of Ced~112, overlaid with our slits showing the extracted apertures. The central, white pixel indicates the position of the star HD~97300. The dash at the bottom left is 10\as.}
\label{Ced112-slits}
\end{figure}

\begin{figure}
\includegraphics[width=3.3in]{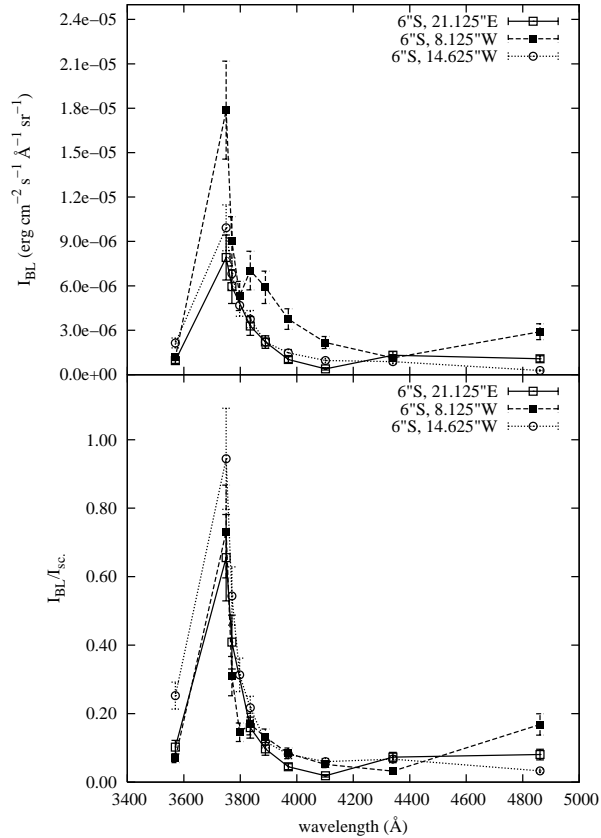}
\caption{BL spectra (upper panel) and ratio of BL to the continuum (lower panel) at 4 different positions in Ced~112. These spectra correspond to extractions along the slit shown in Figure~\ref{ced112ring}.}
\label{Ced112figs}
\end{figure}

We obtained nebular spectra along four slits (PA 90$^\circ$), 0\as\ offset, 6\as\ north, 6\as\ south and 10\as\ north of the central star. Figure~\ref{Ced112-slits} is an optical image of the nebula overlaid with the extracted apertures. We extracted 10 spectra along the 6\as\ north and 6\as\ south slits and 6 spectra along the slit through the center (0\as\ offset) each covering 16.25 sq. arcsec of the nebula. Along the 10\as\ north slit we extracted 4 spectra each covering 32.5 sq. arcsec. BL was measured using the line-depth technique in each of these spectra. Along the 6\as\ south slit we extracted 14 spectra with finer apertures each covering 8.125 sq. arcsec to enable comparison with ISOCAM data discussed in the next section. Figure~\ref{Ced112figs} show representative BL spectra in three such apertures and the ratio of BL to scattered light in the same apertures. 

%place Ced112 figs

\subsubsection{Spatial Correlation of BL with mid-IR emission}
\begin{figure}
\includegraphics[width=3in]{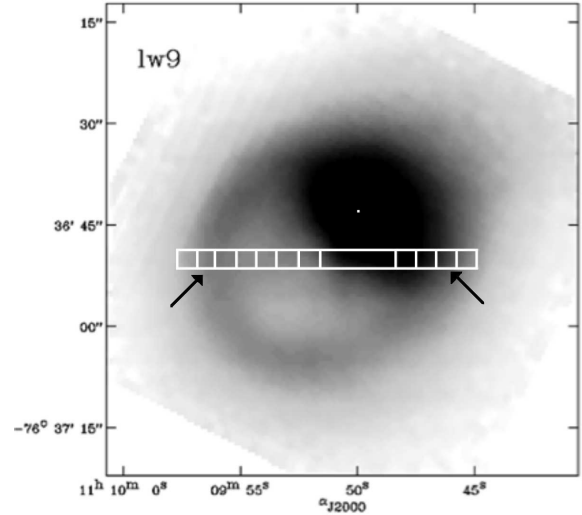}
\caption{ISOCAM image (lw9) of Ced~112 from \citet{siebenmorgen98} overlaid with our slit showing the extraction apertures. The white pixel is overlaid to indicate the position of the central star HD~97300. The arrows indicate where the slit crosses the ring-like structure seen in the image.}
\label{ced112ring}
\end{figure}

\begin{figure}
\includegraphics[width=2.5in,angle=270]{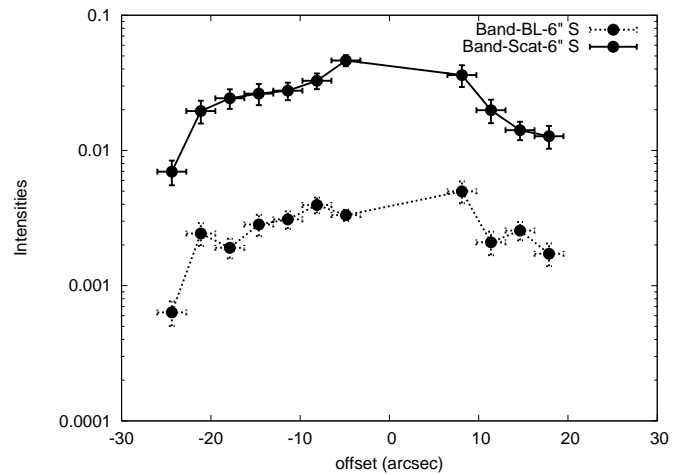}
\caption{Band-integrated BL and scattered light intensities in the apertures indicated in Figure~\ref{ced112ring}. A central-blocker used in the slit prevented measurements close to zero offset. Offsets are along the slit, with zero offset being exactly 6\as\ south of the star.}
\label{ced112dist}
\end{figure}

\begin{figure}
\includegraphics[width=2.5in,angle=270]{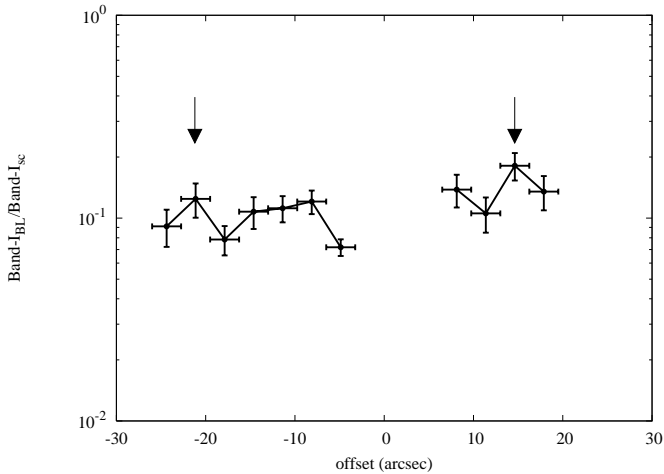}
\caption{Ratio of Band-I$_{BL}$ to Band-I$_{sc}$ in the apertures indicated in Figure~\ref{ced112ring}. Offsets are along the east-west, 6\as\  south slit with zero being exactly south of the star.}
\label{ced112ring-blsc}
\end{figure}

As mentioned earlier, \citet{siebenmorgen98} reported the detection of an extended, ring-like structure around HD~97300 in the mid-IR images of the nebula taken with ISOCAM. They obtained images in four narrow-band filters centered at 6.0  $\mu$m (lw4), 6.8  $\mu$m (lw5), 11.3 $\mu$m (lw8) and 14.9  $\mu$m (lw9). All four images show an extended emission centered on the star and an elliptical ring of size  50\as $\times$36\as\ around it. They also report that the spectra taken with the circular variable filter between 5.8 and 13.8 $\mu$m  roughly aligned along a line intersecting the star, the emission minimum, and the ring seen in the SE direction (PA = 142.4$^\circ$) show strong AEF features. Our 6\as\ south offset slit extends over the nebula and the ring. Figure~\ref{ced112ring} shows the lw9 image overlaid with our slit, with the extraction apertures marked and Figure~\ref{ced112dist} shows a plot of the distribution of the band-integrated BL and dust-scattered light in the apertures along the slit. The integrated BL clearly shows enhancements in the regions where the slit crosses the ring on both sides, while the scattered light does not. It is also interesting to note that in the spectra shown by \citet{siebenmorgen98} the flux at 7.8 $\mu$m between two positions on the ring and outside away from the star varies by a factor of $\sim$~2.4 while the band integrated BL intensity varies by a factor of 2.6 for similar locations. Figure~\ref{ced112ring-blsc} shows the distribution of the ratio of the band-integrated BL to the band-integrated scattered light. The enhancement of the BL at the positions of the ring is clearly seen in this representation as well.

%place Ced112ring and Ced112dist

\subsection{NGC~5367}
NGC~5367 surrounds the double star Her~4636 in the head of the prominent Cometary Globule CG 12 and lies at high galactic latitude (b~=~+21$^\circ$). It is an example of a relatively isolated low-to-intermediate-mass star formation region \citep{vanHill75}, which is thought to have formed as a result of a nearby supernova event 10 -- 20 million years ago \citep{williams77}, and has converted about 20\% of its gas mass into stars \citep{white93}. The most probable distance to NGC~5367 is 630~pc, and the mass of the surrounding molecular cloud is $\sim$~120 M$\sun$\citep{williams77}. \citet{williams77} using optical and infrared photometry showed the likeliest masses of the two stars in the binary system to be 4.5 and 8 M$\sun$, with spectral type B7 and B4, respectively, with the latter being surrounded by a dust-shell, emitting as a 1600~K blackbody. \citet{reipurth93} also obtained spectra of the two stars (called N and S for north and south) as part of their visual pre-main sequence binary program. They show that the northern component has strong H$\alpha$ emission and the southern component, the B7 star, has no such emission, and  they conclude that both are probably Herbig Ae/Be stars. \citet{vieira03} also include Her~4636 in their study of Herbig Ae/Be stars and estimate a rotational velocity of 158 km s$^{-1}$ and an effective temperature of $\sim$~18,700~K.

%place NGC5367-slit
\begin{figure}
\includegraphics[width=3.3in]{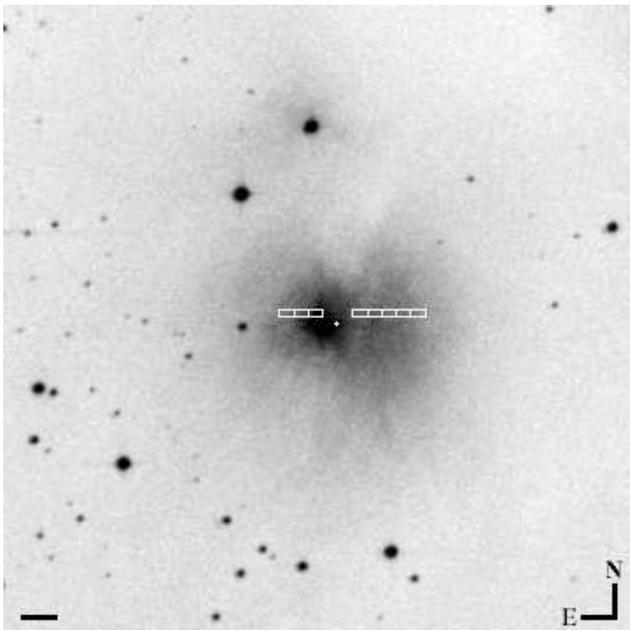}
\caption{AAO.R.DSS2.325 image of NGC~5367, overlaid with our slits showing the extracted apertures. The central, white pixel indicates the position of the star Her~4636. The dash at the bottom left is 20\as.}
\label{NGC5367-slit}
\end{figure}

\begin{figure}
\includegraphics[width=3.3in]{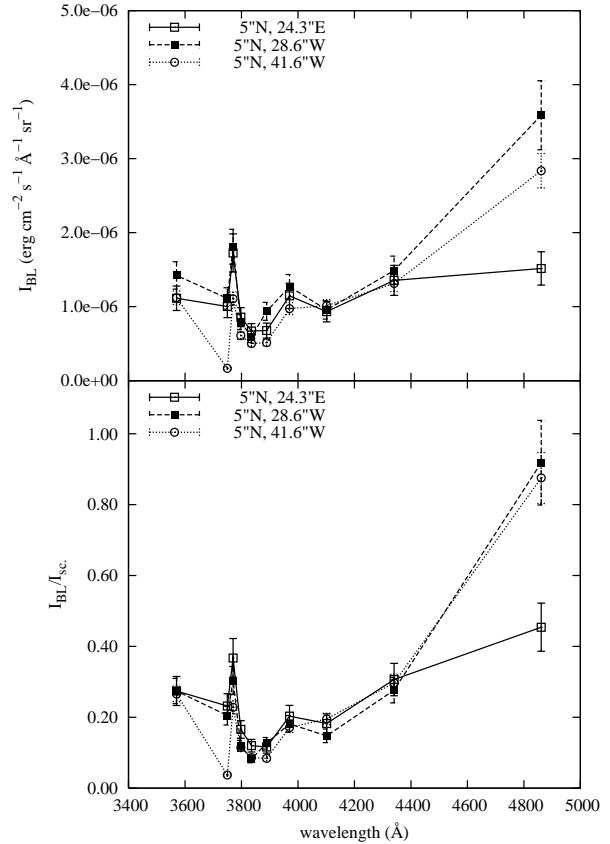}
\caption{BL spectra (upper panel) and ratio of BL to the continuum (lower panel) at three different positions in NGC~5367. These spectra are contaminated with emission from hydrogen recombination, but the peak $\sim$~400~nm, $\sim$~377~nm and the rise towards the Balmer discontinuity at $\lambda$= 357~nm are indicative of BL.} 
\label{NGC5367figs}
\end{figure}

We obtained nebular spectra 5\as\ north of the illuminating stars. Figure~\ref{NGC5367-slit} shows an optical image of the nebula overlaid with the extracted apertures. We extracted 8 spectra each covering 16.25 sq. arcsec of the nebula along the slit. BL was measured using the line-depth technique in each of these apertures. A combined spectrum of both the stars was used for the stellar measurements. Figure~\ref{NGC5367figs} show representative BL spectra in three such apertures and the ratio of BL to scattered light in the same apertures. The red rise in the spectra is a result of emission from hydrogen recombination filling in the hydrogen Balmer lines, in addition to the BL. Characteristic hydrogen recombination spectra alone would result in a spectral intensity steadily decreasing with decreasing wavelength. Not only do these spectra show a peak at $\lambda\sim$~370~nm they also show a secondary peak at $\lambda\sim$~400~nm, which is a feature seen in many BL spectra.

%place NGC 5367 figs

\subsection{NGC~2023}
NGC~2023 is a bright reflection nebula which is a part of the Orion L1630 molecular cloud at a distance of 450 to 500~pc \citep{deBoer83}. This nebula, also known as vdB 52, is illuminated by the bright B1.5 V star HD~37903 \citep{racine68}. NGC~2023 is an example of photo-dissociation regions (PDRs) that exist on the surfaces of molecular clouds in the vicinity of hot, young stars. The interaction of the newborn stars with their parental molecular environment leads to intense emission of fine-structure lines of carbon and oxygen and of H$_2$ ro-vibrational transitions. NGC~2023 has been well studied at many wavelengths, including the mid- and near IR region showing AEFs attributed to PAHs \citep{sel85,ver01,moutou99}. NGC~2023 also exhibits ERE \citep{gr78,wsk84,ws88}.

%place NGC2023-slit
\begin{figure}
\includegraphics[width=3.3in]{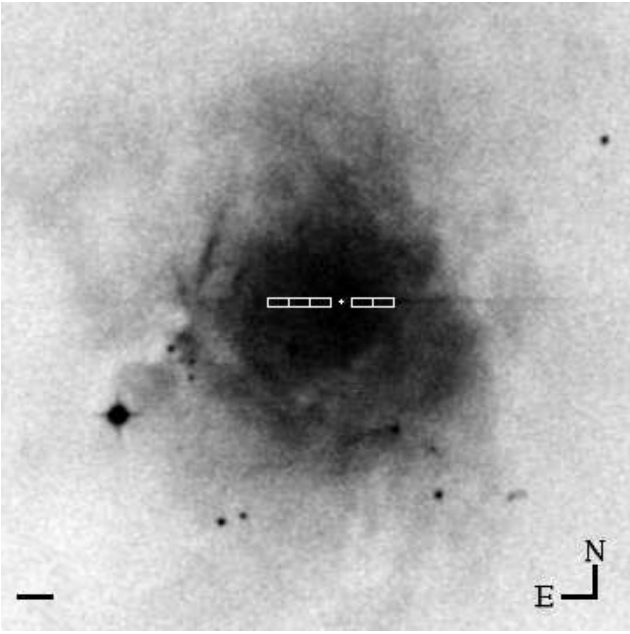}
\caption{SERC.ER.DSS2.768 image of NGC~2023, overlaid with our slits showing the extracted apertures. The central, white pixel indicates the position of the central star HD~37903. The dash at the bottom left is 20\as.}
\label{NGC2023-slit}
\end{figure}

\begin{figure}
\includegraphics[width=3.3in]{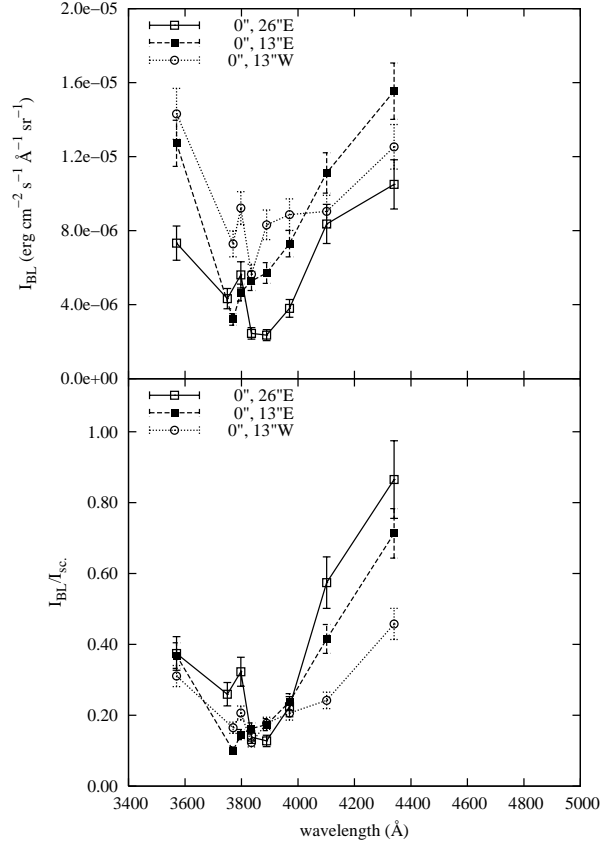}
\caption{BL spectra (upper panel) and ratio of BL to the continuum (lower panel) at three different positions in NGC~2023. These spectra are heavily contaminated with emission from hydrogen recombination, but the peak $\sim$~390~nm and the rise towards the Balmer discontinuity at $\lambda$= 357~nm are indicative of BL.}
\label{NGC2023figs}
\end{figure}

We obtained nebular spectra with a slit (PA 90$^\circ$) at 0\as\ offset with a central decker blocking out the star. Figure~\ref{NGC2023-slit} shows an optical image of the nebula overlaid with the extracted apertures. We also obtained spectra of the star HD~37903. We extracted 5 spectra each covering 32.5 sq. arcsec of the nebula along the slit. BL was measured using the line-depth technique in each of these apertures. Figure~\ref{NGC2023figs} show representative BL spectra in three such apertures and the ratio of BL to scattered light in the same apertures. As in NGC~5367, the red rise in the spectra is a result of hydrogen recombination filling in the hydrogen Balmer lines in addition to the BL. In this case the origin of this recombination emission is the rather extended region of photo-ionized hydrogen gas overlying the Orion constellation. There is no measurement of BL at H$\beta$ because the nebular H$\beta$ ($\lambda \sim$~4861~\AA) emission more than fills the scattered stellar H$\beta$ absorption line. Characteristic recombination spectra alone would result in a spectrum steadily decreasing with decreasing wavelength. These spectra show a upturn at $\lambda\sim$~400~nm and a pronounced peak at $\lambda\sim$~380~nm, which we interpret as evidence for the presence of BL.

%place NGC2023 figs

\section{DISCUSSION}
We have presented optical long-slit spectra of reflection nebulae illuminated by stars having a range of temperatures between 10,000 -- 23,000~K, and thus covering a wide range of radiation environments. The long-slit technique enables us to extract many individual spectra in each nebula that differ in local radiation density and probably size-distribution of emission carriers but do not differ considerably in chemical composition or grain-processing histories. Comparison of BL spectra from different nebulae will help us study the evolution of the carriers with changing environments and also see what effects different grain processing has on the BL emitters.

\subsection{\label{sec-spec}The BL Spectrum}
\begin{figure}
\includegraphics[width=2.5in,angle=270]{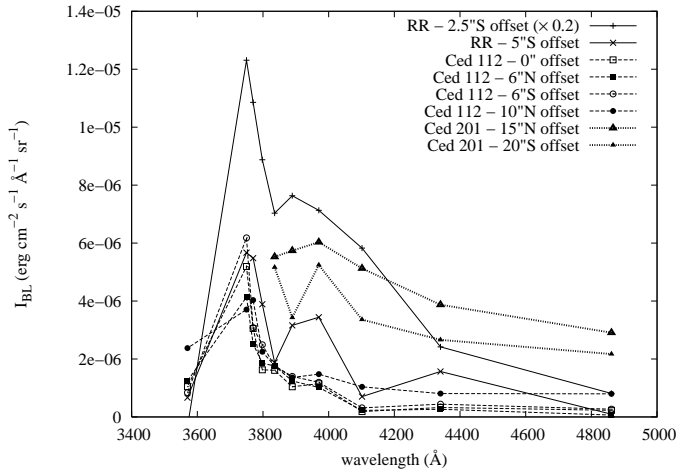}
\caption{Average BL spectra from the four slits in Ced~112, and two slits in Ced~201 compared to two spectra from the RR.}
\label{NormBL}
\end{figure}

\begin{figure}
\includegraphics[width=2.5in,angle=270]{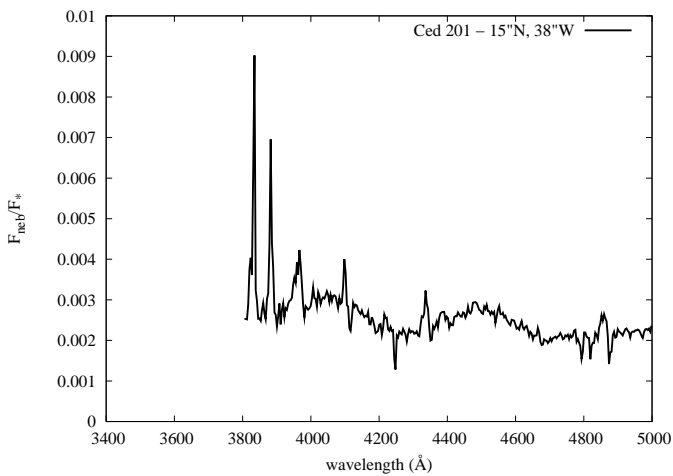}
\caption{Ratio of the nebular spectrum to the stellar spectrum in Ced~201 from a position 15\as\ north and 38\as\ west of the central star. The two broad bands at $\lambda\sim$~4050~\AA and $\lambda \sim$~4500~\AA\ are new features and are likely to be signatures slightly larger PAHs ($>$ 14 C atoms).}
\label{Ced201-band}
\end{figure}

The BL spectra in different apertures along the slit in the same nebula varied in intensity with offset from the respective stars but were fairly similar in spectral shape, and therefore we produced one average spectrum for each of the slits. This allows us to identify differences between the BL spectrum in different nebulae and find possible correlations to varying environments and local ISM histories. Figure~\ref{NormBL} shows the average BL spectrum from the two slits in Ced~201, four slits in Ced~112 and two BL spectra from the RR (reported earlier in \citet{vijh04}). In this section we only discuss Ced~201 and Ced~112. The other two nebulae NGC~2023  and NGC~5367 will be discussed in a later section (\S~\ref{sec-Hion}). The RR BL spectrum from the slit 2$\farcs$5 south offset has been scaled down by a factor of 5 to aid comparison. Four differences in these spectra may be noted. One, the BL intensity in the RR-2$\farcs$5 south is about 10 times higher than in Ced~112 but the RR-5\as\ south is similar in intensity. Two, the RR spectra have peaks at $\lambda\sim$~377~nm similar to Ced~112. Three, the RR spectra have a long-wavelength component that is absent in Ced~112, but present in Ced~201. And finally, the Ced~201 BL spectrum is unlike that of Ced~112, but could be similar to that of the RR, given the short wavelength rise, but lack of data short-ward of 380~nm makes it difficult to classify the spectrum. All the spectra also show a secondary peak at $\lambda\sim$~400~nm, which could be an indication of a population of efficient fluorescing molecules of slightly larger sizes. The nebular spectra in Ced~201 show some interesting band-like emission features, which are not present in the stellar spectrum and therefore must be nebular in origin. Figure~\ref{Ced201-band} shows the nebular spectrum at 15\as\ north, 38\as\ west divided by the stellar spectrum. This resulting spectrum is free of any stellar features and the peaks at the positions of the Balmer lines indicate filling in of the nebular lines by the BL. The two broad bands at $\lambda\sim$~4050~\AA and $\lambda \sim$~4500~\AA\ are new features and are likely to be signatures of slightly larger PAHs ($>$ 14 C atoms) than the 3-4 ringed ones that produce the blue peak at $\lambda\sim$~377~nm.  

%place NormBL and Ced201-band

\citet{kemper99} model the far-IR and sub-millimeter observations of Ced~201 and conclude that heating by the ionization of PAHs (carbonaceous particles with linear size $>$ 15~\AA) is required to explain the observed properties of Ced~201. Creation of PAHs that produce the observed mid-IR features  by processing of small carbonaceous grains is also invoked in this nebula \citep{cesarsky00}. \citet{cesarsky00} suggest that some of carbonaceous material that emits the mid-IR continuum and broad bands far from the star is processed through the effect of the star that moves through the molecular cloud, producing AEF carriers: the continuum is only 2 times fainter 12\as\ from the star than close to the star, demonstrating the partial disappearance of its carriers near the star while the AEFs become very strong close to the star. They also infer that these carbonaceous grains must be very small (radius of the order of 1~nm) since they are heated transiently by visible photons to the temperatures of $\simeq$~250~K necessary to emit the observed mid-IR features. It is also interesting to note that \citet{witt87} find from UV/visible scattering studies of Ced~201 that the grains responsible for the visible scattering have a narrow size distribution skewed towards big, wavelength - sized grains. This is mostly seen within about 20\as\ from the star. On the other hand the exceptionally strong 2175~\AA\ extinction band observed in the spectrum of BD~+69~1231, the illuminating star, suggests the presence of many smaller carbonaceous grains. All this is evidence for grain processing by the radiation of the star, or/and by shattering of grains by a shock wave associated with its motion. The banded spectrum (Figure~\ref{Ced201-band}) is of the arc-like structure seen toward the NW in the DSS image of the nebula shown in Figure~\ref{Ced201-slit}. This arc is probably a result of a shock and we suggest that the BL carriers are produced in situ from the larger AEF carriers being photo-dissociated in the stellar UV-field. 

\subsection{Spatial Distribution of BL}
\begin{figure}
\includegraphics[width=3.3in]{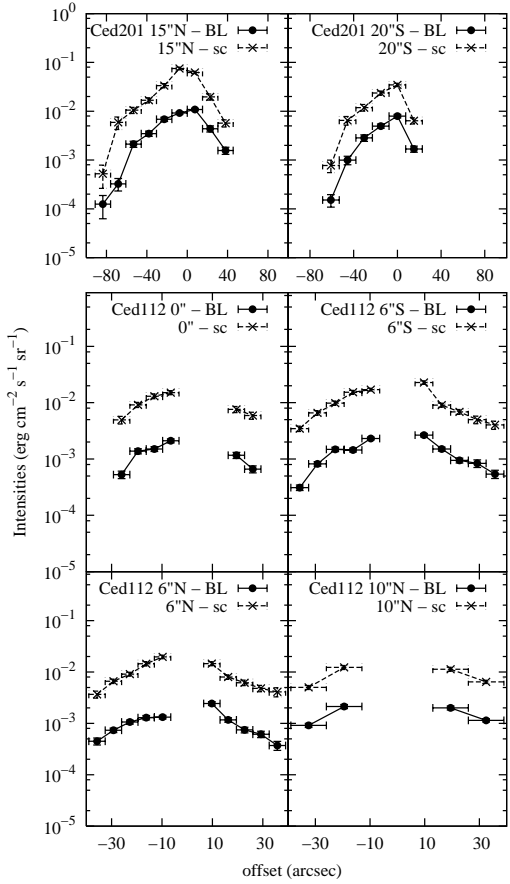}
\caption{Band-integrated BL and scattered light intensities along the various slits in Ced~201 (upper panel) and Ced~112 (middle and lower panel).}
\label{band-bl}
\end{figure}

We observed little variation in spectral shape in the BL spectra extracted from different apertures along a given slit, in all the nebulae, but considerable variation in the intensity of both the BL and the scattered light. We plot band-integrated BL and scattered light intensities as a function of offset from the central illuminating star for each slit position. We integrate from H$\beta$ to the Balmer discontinuity (486~nm -- 357~nm) for both the BL and the scattered light in case of Ced~112 and till 383.5~nm at the lower wavelength limit for Ced~201 because of the wavelength filter that was used for those observations. Figure~\ref{band-bl} shows band-integrated BL and scattered light intensities as a function of offset from the central star along the six slits in the two nebulae. The distribution of BL in Ced~112 seems to be flatter than that in Ced~201, but the overall BL-intensities in Ced~201 are about five times higher than those in Ced~112. Compared to the maximum intensity of the BL in the RR (2$\farcs$5 south slit) the intensities in Ced~112 and Ced~201 are 20 times and 4 times lower respectively. 

%place band-bl and blDIVsc

\begin{figure}
\includegraphics[width=3.3in]{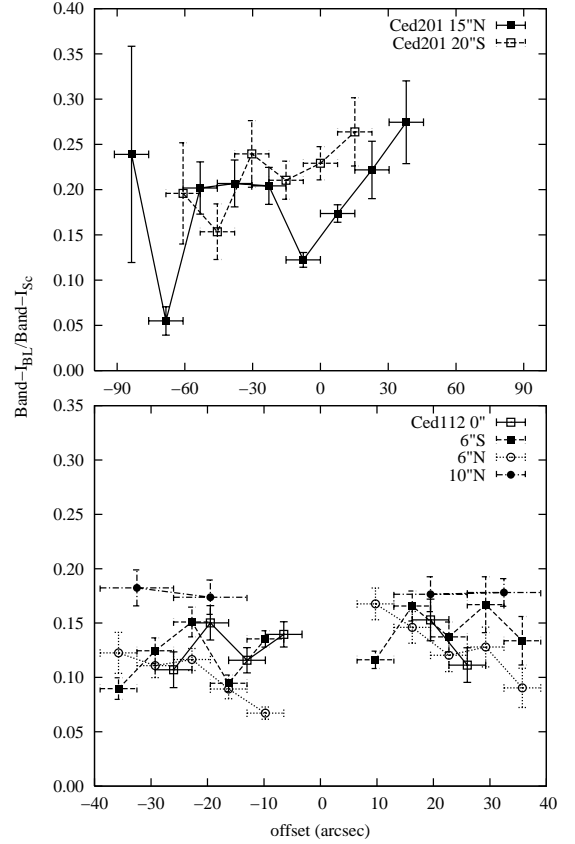}
\caption{Ratio of the band-integrated BL to band-integrated scattered light in Ced~201 (top panel) and Ced~112 (bottom panel). On large scales the BL distribution is uniform compared to the scattered light in Ced~112 but shows asymmetries in Ced~201.}
\label{blDIVsc}
\end{figure}

Our BL observations of Ced~112 also trace out the ring of of AEF emitters seen in mid-IR observations by \citet{siebenmorgen98}. However, taking the complete distribution over all four slits into account, the large-scale distribution of both BL and scattered light is fairly uniform. This is also borne out by the distribution of the ratio of the band-integrated BL to the scattered light intensity shown in Figure~\ref{blDIVsc} (bottom panel). The ratio of BL to scattered light remains fairly constant over large areas of the nebula.

The spatial distribution of the BL in Ced~201 is interesting. The BL seems asymmetrically distributed with respect to the star; this is primarily due to the asymmetry in the nebulosity itself. Figure~\ref{Ced201-slit} shows that the nebula is much more extended to the east than the west, toward the molecular cloud that the star is ploughing into. BL can be detected up to 80\as\ into the cloud, though the intensity falls off rapidly. On the west the BL and the scattered light both fall off more quickly, but where the 15\as\ north slit crosses the arc-like shock-structure, the nebular spectrum shows the distinct banded features discussed earlier in \S~\ref{sec-spec}.This complex distribution is true of other tracers as well. The CO(2--1) map of the molecular cloud is distinctly asymmetric with double peaked profiles to the north and single peaked profiles to the south whereas the AEF spectra on the other hand are remarkably symmetric \citep{cesarsky00}. The relative intensity of BL to the scattered light is shown in the upper two panels of Figure~\ref{blDIVsc}. The forward-directed scattered light falls of much more rapidly than the more isotropic BL, and thus the ratio rises at larger offsets.

\subsection{\label{sec-Hion}BL in Hydrogen Ionization Regions}
\begin{figure}
\includegraphics[width=2.5in,angle=270]{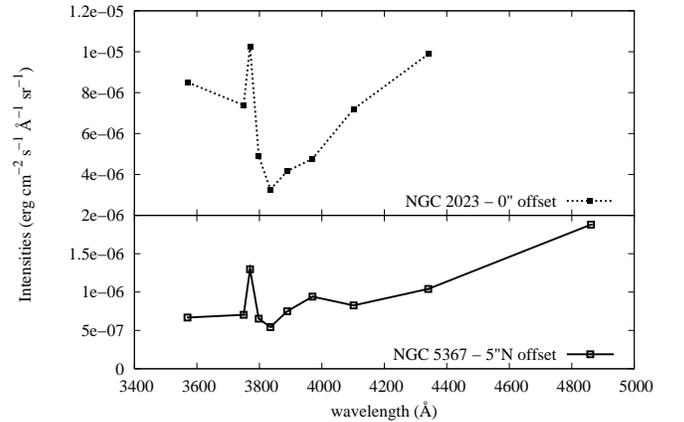}
\caption{Average BL spectrum in NGC~2023 (top panel) and NGC~5367 (bottom panel). Contaminated by emission from hydrogen recombination these are only marginal detections.}
\label{2023-5367-avgBL}
\end{figure}

NGC~5367 and NGC~2023 are both nebulae where the detected spectrum is heavily contaminated by emission from hydrogen recombination. Again as the individual spectra in each slit do not vary in spectral character we present an average BL spectrum for each of the slits in the two nebulae in Figure~\ref{2023-5367-avgBL}. The two show similar red spectra with intensities decreasing towards the higher-order Balmer lines characteristic of recombination emission from ionized hydrogen but with peaks slightly short-ward of $\lambda\sim$~380~nm. The NGC~2023 spectrum is more severely contaminated than NGC~5367 because of a stronger ionizing radiation field, and the NGC~5367 spectrum also shows a secondary peak at $\lambda\sim$~400~nm similar to other BL spectra, in addition to the peak short-ward of 380~nm. The measurement at the Balmer-jump ($\lambda$~=~357~nm) in the other BL spectra is a measure of difference in absorption, but in these spectra it is a measure of emission from ionized hydrogen. Nevertheless, the peak in the spectrum at $\lambda <$~380~nm and the secondary feature at 400~nm are signatures of BL, albeit marginal. 

%place 2023-5367-avgBL

Observation of extended emission, especially in the blue region of the spectrum is challenging because of competing contribution from scattered light, and emission from hydrogen recombination succeeds in swamping it even further. We would advise other observers to avoid regions with ionized hydrogen in future searches of BL.

\subsection{\label{sec-form}Survival vs. In-situ formation of BL Carriers}
BL was first discovered in the RR, a proto-planetary nebula surrounding the evolved post-AGB star HD~44179. This is a region of active dust production, and therefore it is not surprising to see BL from small molecules with high fluorescence efficiencies like anathracene or pyrene, provided there are environments where such species can remain neutral. The regions in the shadow of the circum-binary disk of the central binary provides just such a sheltered environment, and this is where the BL in the RR is predominantly observed. Therefore, a ``bottom-up'' production mechanism where the PAH molecules are built from smaller carbon structures is feasible in the RR.  Such small PAH molecules are not expected to survive in the ISM, being photo-dissociated under interstellar radiation conditions \citep[][and references therein]{jbl99,lepage03}. Therefore we did not expect to see BL in other nebulae and predicted it's absence in \citet{vijh05}. However, this prediction is valid only if the only possible means for the existence of these molecules in the other nebulae was travel through the ISM: time-scales of such travel are much longer than the expected molecular lifetimes in such environments. 

Therefore, the detection of BL in other, non-carbon-rich nebulae came as a surprise. However, the formation of PAH molecules through a ``top-down'' process emerges as a likely alternative formation mechanism of small PAHs in these locations. The presence of hydrogenated amorphous carbon (HAC) solids in interstellar space is well established through the observation of the interstellar absorption feature at 3.4~$\micron$ \citep[see][for a recent review]{pendleton04}. In an important laboratory investigation, \citet{scott96} demonstrated that heating of HAC solids will lead to the evolution of aromatic structures, distinguished by their characteristic 3.29~$\micron$ C-H stretch feature. Subsequently, \citet{scott97} showed that HAC solids will decompose under intense UV radiation into an extended sequence of aromatic carbon clusters with a wide mass range, with small molecules ($<$~20 atoms) appearing first under the lowest-fluence conditions. These laboratory investigations followed earlier theoretical work by \citet{duley92} and \citet{taylor93}, suggesting that even smaller hydrocarbons, in particular CH$^+$, could be produced through the erosion of carbonaceous dust in shocks, cloud-intercloud interfaces and PDRs. We envision that a similar process could be responsible for the appearance of small PAH molecules in reflection nebulae. Furthermore, larger PAH molecules that are able to survive conditions in the diffuse ISM are likely to undergo photo-fragmentation, as recently suggested by \citet{rapacioli05} to explain the variation in the sizes of of AEF emitters across a PDR in the reflection nebula NGC~7023.  	

We conclude, therefore, that the most likely source of the small PAHs in reflection nebulae, detected through the BL, are larger carbonaceous structures such as HAC grains or large, stable PAH clusters. The likely cause is a process of photo-dissociation and photo-fragmentation, driven by the strongly enhanced radiation fields of the nearby early-type illuminating stars. There is other ancillary evidence for grain processing in these environments, in particular in Ced~201 \citep{cesarsky00}. New molecules are being formed as the star BD~+69~1231 rams into the molecular cloud, exposing previously sheltered environments to UV radiation. In this case, we are witnessing a dynamic process where these molecules are both being created from larger species and destroyed by impinging UV radiation. In answer to the question posed in the introduction: \emph{Do small PAHs survive in the ISM? or are the AEFs primarily produced by larger PAHs?}: Yes, small PAHs survive as parts of larger entities and are liberated again in environments of enhanced radiation density. They do exist in regions where AEFs are observed and should be taken into account in models that try to explain the AEFs. However, since BL is strongly biased toward small PAH molecules, the detection of a BL signal does not preclude the presence of a much wider distribution of sizes of aromatic structures. 

\section{CONCLUSIONS}
We have presented here BL spectra in reflection nebulae illuminated by stars having temperatures between 10,000 -- 23,000~K. Of the nebulae studied, Ced~201 and NGC~2023 have previously been shown to exhibit ERE as well \citep{wb90,wsk84,ws88}. ERE has also been seen in the other two nebulae NGC~5367 and Ced~112 (U. Vijh et al., in preparation). Many of these nebulae also show the AEFs, attributed to PAHs. Our BL spectra which we have previously identified as fluorescence from small, neutral PAHs provide evidence that these small molecules are present in the ISM. Given that these small PAHs (with less than 20 carbon atoms) are not expected to survive harsh interstellar conditions, we believe that the BL spectra provide evidence for the production of these small molecules from larger structures by the impinging UV radiation from the central stars. 

\acknowledgements
This research has been supported by the National Science Foundation through Grant AST 0307307 to The University of Toledo. This research has made use of NASA's Astrophysics Data System Bibliographic Services, the SIMBAD database, operated at CDS, Strasbourg, France and also of Aladin image server.

\bibliography{thesis_ref}

\clearpage

\end{document}